# The Triad of Modern Democracies: Money, Identity, and Information in Shaping Power and Legitimacy

Venkat Ram Reddy Ganuthula[1] Krishna Kumar Balaraman[1]

[1] Indian Institute of Technology Jodhpur

## Abstract

This article examines the interplay of money, identity, and information as a pivotal triad reshaping electoral politics and legitimacy in modern democracies, with insights from the United States, India, Germany, China, and Russia. Financial resources, through campaign finance and state funds, enable strategies exploiting identity cleavages like race, caste, and nationalism, amplified by digital networks such as social media and targeted messaging. In democracies, this dynamic fosters polarization and erodes trust, while in non-democracies, it bolsters regime narratives. Drawing on political economy, social identity theory, and media studies, the study reveals a feedback loop: money shapes identity appeals, information disseminates them, and power consolidates, challenging issue-based governance assumptions. Comparative analysis highlights the triad's universal yet context-specific impact, underscoring the need for reforms to address its effects on democratic theory and practice, as it entrenches elite influence and tribal divisions across diverse political systems.

## 1. Introduction

Contemporary democracies are at a crossroads, marked by forces that undermine the orderly progress once predicted by political behavior scholars. At the heart of this reality is a three-way mix of money, identity, and information---connected pillars that capture what can be called the "new age" of democratic governance. Money, as a capital asset, drives political campaigns, sets policy agendas, and skews power in the direction of those who own it, from deep-pocket donors in open electoral systems to state treasuries in controlled ones. Its political role, long seen as a tool of influence, has been more salient in an age of deregulated campaign finance and state-controlled economies. Combined campaign spending in the 2024 United States federal election totaled an estimated $16 billion, including both presidential and congressional contests, including super PACs and outside spending organizations, with presidential candidates raising $1.6 billion and congressional candidates $3.3 billion as of September 30, 2024, a surge from the $14.4 billion spent in 2020 (OpenSecrets 2024; Federal Election Commission 2025). In democracies, it gives elites the resources to shape elections and policy, often drowning out the voices of the ordinary citizen (Gilens and Page 2014). In non-democracies, it maintains patronage networks that ensure loyalty and repress dissent (Treisman 2020).

Identity, a persistent aspect of human social life, ties citizens to parties and leaders to deep attachments---race, religion, caste, or region---based on evolutionary requirements for group solidarity. New evidence attests to it: in the United States, 83% of Black voters voted for Kamala Harris in 2024, while 42% of Latinos voted for Donald Trump, as evidence of the power of racial identity (PBS News 2024). In India, Muslims have coalesced against the BJP, with 77% voting for the Mahagathbandhan in Bihar (2020) and 79% for the Samajwadi Party in Uttar Pradesh (2022) (Ahmad 2024). Information, driven by the digital revolution, amplifies these trends, mobilizing money power and identity appeals into the public arena with previously unprecedented speed and precision. Information technology has frequently served to amplify polarization, facilitating micro-targeted messaging that serves to harden "tribal" boundaries of belonging, creating echo chambers that amplify societal cleavages (Benkler et al. 2018; Vaidhyanathan 2018; Kubina and von Sikorski 2021). In combination, these forces form a feedback loop that entrenches political power, frequently replacing the rational, issue-based deliberation long celebrated as democracy's promise. This perspective article examines how this triad functions in various political systems---democracies like the United States, India, and Germany, and non-democracies like China and Russia---demonstrating its ubiquity in structuring governance and undermining modernist assumptions regarding political progress.

The assumption that democracies would become systems where voters prioritize policy platforms over local loyalties is deeply embedded in the political science tradition. Downs' economic theory of democracy characterized voters as utility maximizers, voting for candidates who best represented their individual material self-interest (Downs 1957). The later rational choice model, elaborated by theorists like Enelow and Hinich, assumed that electional competition would be a spatial question of proximity to voter opinion, with parties clustering around the median voter in two-party systems or splintering in multi-party systems (Enelow and Hinich 1984). Modernization theorists based their theory on this rosy expectation, predicting that economic growth, expansion of education, and technological change would reduce identity-based voting, with more attention to issue-based politics. Lipset argued that industrialization and literacy would produce a politics of class and economic interests (Lipset 1960), while Inglehart predicted that post-materialist values---emerging in wealthy societies---would shift attention away from ascriptive affiliations towards concerns about quality of life (Inglehart 1997). Theoretical models predicted a linear process: as societies developed, democracy would be a marketplace of ideas, not a war zone for tribal identities.

But behind this narrative lies the resilience of money, identity, and information as dominant forces. Rather than dissipating, these have co-evolved with modernity, entangling each other to structure elections and legitimacy in ways that resist the anticipated decline. In democracies such as the United States, campaign finance has skewed policy towards elite interests, eroding public trust in electoral fairness (Gilens and Page 2014). In India, financing mechanisms such as electoral bonds, valued at INR 16,518.11 crore from 2018 to 2024, with the BJP reaching 47.5%, have funded ruling parties at the expense of transparency and accountability (Election Commission of India 2024; The Hindu 2024). In non-democracies such as China and Russia, state-hold wealth---either through firms or rents on natural resources---secures regime stability, albeit one that masks the vulnerabilities revealed by economic transformation or internal purge (Naughton 2018; Sakwa 2009). These examples suggest money's dual nature: a visible distorting force in open systems, a hidden stabilizer in closed systems. Parallel observations on identity politics suggest its abiding relevance to democratic voting, resisting pressure of modernization towards issue-based competition (Achen and Bartels 2016; Mason 2018). Racial alignments in the United States, caste and religious blocs in India, and regional cleavages in Germany illustrate how identity structures political action, often amplified by digital tools rather than diminished (Tesler 2016; Chhibber and Verma 2018; Kumar and Singh 2014; Mader and Schoen 2021).

The commonality of these observations lies in the acknowledgment that money and identity, each strong alone, derive further strength from being paired together---a pairing made possible through information. In the United States, campaign funding focuses on racial and cultural identity, effectively segmenting the electorate through advanced digital technology, with the likes of Facebook projecting these cleavages (Sides et al. 2018; NYU Stern 2021). In India, monetary donations reinforce caste and religious appeals, disseminated through ground-level messaging applications like WhatsApp (Meta 2024; Kumar and Singh 2014). Even in Germany, a managed political terrain, funding underlies parties representing regional identity, further supported by media narratives (Mader and Schoen 2021). Authoritarian regimes present a variation: in China, state-controlled enterprises link economic growth with nationalist identity, crafted through managed information outreach, while in Russia, resource abundance supports a patriotic narrative, maintained through propaganda (Naughton 2018; Levada Center 2023). Throughout these varied environments, the medium is information---whether through social media in democratic societies or state-controlled media in authoritarian regimes---bringing money and identity together under a single triad that determines political outcomes.

This exchange challenges fundamental assumptions about the progress of democracy. The ground of modernization theory---that development would make political participation superior to tribal allegiances---erodes as identity continues to be relevant, fueled by economic resources and knowledge (Inglehart 1997). Rational choice models, which expect that citizens would vote based on policies rather than loyalty, face challenges in explaining the resilience of partisan and group loyalties (Downs 1957; Enelow and Hinich 1984). Techno-optimism, anticipating that online spaces would facilitate discussion across divides, has shifted to an acceptance of their role in fueling polarization (Benkler 2006; Sunstein 2017; Barrett et al. 2021). The dominant power of this triad suggests that democracy, rather than rising above these components, accommodates them, raising questions about its capacity to deliver ideals of representation and fairness. In democratic environments, the transparency of this process erodes confidence as citizens witness elite manipulation and tribal cleavages; in non-democratic environments, the obscurity hides vulnerabilities that can surface in times of crises, from economic downturns to elite defectors.

The comparative reach of this research---within the United States, India, Germany, China, and Russia---offers the context to consider these differences. The United States is an example of an open system in which financial resources and information support partisan and racial identities, too often at the expense of legitimacy. India, in its dynamic multi-party democracy, is an example of how financial power manipulates religious and caste identities, amplified by digital networks, though moderated by judicial checks. Germany, with its regulated democratic system, is a balanced example, with funding and media supporting regional identities within a stable system. China and Russia, as non-democracies, are examples of how state-controlled financial resources and information support regime-aligned identities, asserting authority while concealing potential threats. These examples, drawn from a range of institutional and cultural contexts, address the universality of the triad and its context-dependent expressions, across experiences in democratic and authoritarian regimes.

This perspective article elaborates on these observations to argue that money, identity, and information constitute a primal triad in contemporary democracies with implications that reach beyond electoral politics to the character of political legitimacy. It argues that the triad, far from representing a distortion, is an intrinsic aspect of modern governance---one that resists linear models of democratic development and necessitates a rethinking of theory and practice. The analysis runs over a number of sections. A theoretical framework draws on political economy, social identity theory, and media studies to position the dynamics of the triad. The next sections examine each component---money, identity, and information---in isolation, drawing on empirical examples to identify their respective roles. A comparative synthesis then examines their interlinks, comparing democratic and non-democratic regimes. The last section discusses the triad's negative effects and suggests avenues for reform and theoretical enhancement.

By putting money, identity, and information front and center, this article aims to enrich the debates about the future of democracy. It transcends the diagnosis of discrete forces to shed light on their synergy, providing a multifaceted map of power concentration in an age of technological and social complexity. The resilience of the triad implies that good government does not demand the suppression of these elements---an impossibility in a utopian fantasy---but their regulation through policy ingenuity and institutional handcraft. As democracies face polarization, lack of trust, and global pressure, an appreciation of this triad provides a point of departure for reconceiving political stability and legitimacy in the twenty-first century.

## 2. Theoretical Framework

The interaction of money, identity, and information provides a critical perspective with which to analyze the dynamics of contemporary democracies, in which electoral practices and the perception of political legitimacy deviate from the rationalist axioms defined by classical theory. This section formulates a theoretical model to account for their interactions, invoking the expertise of political economy, social identity theory, and media studies to account for how the three factors shape influence in various systems—democracies such as the United States, India, and Germany, and non-democratic systems such as China and Russia. Money is a structural asset that facilitates influence by funding campaigns and patrons; identity is a psychological and evolutionary anchor that ties voters to political actors; and information is a technological medium that enhances these effects in situations of digital interconnectivity. Together, these elements form a feedback loop that enhances authority, thereby undermining modernist expectations for governance on specific issues. This model synthesizes the literature to contend that the triad is a necessary aspect of contemporary politics, necessitating a reorientation of democratic theory.

Political economy provides the foundation for understanding money's role in the triad. Scholars like Gilens and Page have demonstrated that economic resources skew policy toward elite interests in democracies, a pattern rooted in the ability of wealth to dominate electoral processes, as seen in the U.S.'s $16 billion 2024 election spending (Gilens and Page 2014; OpenSecrets 2024). For example, Gilens and Page (2014) find that U.S. government policy aligns with the preferences of affluent groups far more than those of average citizens, illustrating how money can drown out ordinary voices. This perspective builds on elite theory, where Mills argued that power concentrates among a small cadre of economic and political actors who leverage resources to maintain dominance (Mills 1956). In open systems, money flows through legal channels—campaign contributions, lobbying, and party funding—amplifying the voices of those who can pay, whether corporations in the United States or industrialists in India, where electoral bonds totaled INR 16,518.11 crore, with the BJP receiving nearly half (The Hindu 2024). Corporate and billionaire donors in the United States exert outsized influence on candidate selection and agendas (Bartels 2016). In closed systems, such as China and Russia, state-controlled wealth—via enterprises or resource rents—serves a similar function, securing loyalty and suppressing dissent (Naughton 2018; Treisman 2020). In China, the Communist Party's control over vast state enterprises and budgets allows it to reward cadres and co-opt potential opposition, blurring state and party finances (Ang 2020). In Russia, oil and gas revenues fill state coffers that President Putin deploys to enrich loyal oligarchs and regional bosses, creating a vertical of payments that entrenches his rule (Gurvich and Prilepskiy 2015). Political economy thus frames money as a

mechanism of influence, not merely a neutral tool, shaping governance by prioritizing those with access to it over the broader polity.

This structural role of money intersects with electoral competition, where rational choice models once held sway. Downs posited that voters select candidates based on policy proximity, with parties adjusting positions to capture the median voter (Downs 1957). Enelow and Hinich refined this spatial theory, suggesting multi-dimensional issue spaces where parties compete for voter support (Enelow and Hinich 1984). Yet, these models falter when money distorts the playing field. In democracies, financial backing enables campaigns to drown out smaller voices, as seen in the United States' deregulated system or India's funding mechanisms (Federal Election Commission 2025; Kapur and Vaishnav 2018). However, these models struggle to account for what we observe in practice: elections often turn on identity loyalties and marketing battles, not just policy positions. Voters do not behave as purely rational utility-maximizers; instead, they are influenced by partisan team spirit and social identities (Achen and Bartels 2016). In non-democracies, state resources ensure a monopoly on power, sidelining opposition (Gurvich and Prilepskiy 2015). Money, then, does not facilitate a fair marketplace of ideas but tilts it toward those who can afford to dominate, undermining the rationalist assumption of equal voter influence.

Identity, the second pillar, shifts the focus from economic incentives to social-psychological drivers, rooted in human evolution. Social identity theory, developed by Tajfel and Turner, argues that individuals derive self-concept from group memberships, fostering in-group loyalty and out-group differentiation (Tajfel and Turner 1979). Brewer extends this to politics, suggesting that the need for positive distinctiveness drives group-based behavior, a trait honed by millennia of tribal survival (Brewer 1999). In politics, this means many voters support the party that represents their group—racial, ethnic, religious, or regional—regardless of specific policy details. Brewer's (1999) work on in-group loyalty suggests that such attachments can outweigh policy considerations: people feel "us versus them" instincts strongly in political contexts. In democracies, this manifests as stable partisan or ethnic alignments—racial blocs in the United States, where 83% of Black voters backed Harris in 2024; caste and religious affiliations in India, with 79% of Muslims supporting anti-BJP parties in Uttar Pradesh (2022); regional divides in Germany, where civic identity boosts turnout (Achen and Bartels 2016; Chandra 2004; Arzheimer and Berning 2019; PBS News 2024; Ahmad 2024; Mader and Schoen 2021). In non-democracies, regimes harness identity—nationalism in China, patriotism in Russia—to unify support and legitimize control (Naughton 2018; Sakwa 2009). Identity's power lies in its emotional resonance, transcending the calculated self-interest of rational choice models. Identity, then, is not an irrational residue to be erased by modernity, but an adaptive force in democratic politics that parties and leaders actively mobilize.

This persistence challenges modernization theory's optimism. Lipset predicted that economic development would shift politics toward class-based interests, eroding ascriptive ties (Lipset 1960). Inglehart went further, suggesting that post-materialist values would supplant identity with issue-driven concerns in advanced societies (Inglehart 1997). Yet, identity endures, adapting to modern contexts rather than fading. In the United States, racial attitudes outpredict economic factors in voter choice (Tesler 2016); in India, caste and religion anchor party systems despite economic development. Non-democracies like China and Russia exploit identity—Han Chinese pride or Slavic unity—to sustain legitimacy, often through state-controlled narratives (Naughton 2018; Levada Center 2023). Far from a pre-modern relic, identity thrives as a constitutive force, amplified by the triad's other elements.

The Michigan model of partisanship complements this view, framing party loyalty as a social identity rather than a policy tally (Campbell et al. 1960). Achen and Bartels extend this, arguing that voters align with groups reflecting their symbolic attachments—race, religion, region—over material interests (Achen and Bartels 2016). This "group interest" lens reveals why policy shifts rarely disrupt allegiances: voters adjust beliefs to fit their identity, not vice versa (Lenz 2012). In democracies, this explains the durability of partisan divides; in non-democracies, it underpins regime support despite economic fluctuations. Identity, then, is not irrational but a heuristic, simplifying complex political choices in ways rational choice models overlook (Huddy 2001). Its evolutionary roots—group survival through cooperation—make it a resilient anchor, exploited by those wielding money and information.

Information, the third pillar, bridges money and identity through its role as a technological amplifier. Early optimists like Benkler foresaw a networked public sphere where digital tools would democratize discourse, enabling deliberation across divides (Benkler 2006). Shirky envisioned collective action driven by shared interests, not inherited loyalties (Shirky 2010). Yet, reality has diverged sharply. Digital platforms—social media, messaging apps, algorithms—intensify the triad's effects, channeling money into identity-based mobilization and entrenching divisions (Vaidhyanathan 2018; Sunstein 2017). In democracies, targeted campaigns amplify racial, caste, or regional appeals, as seen in the United States' partisan echo chambers or India's grassroots digital networks (Barberá 2015; Kumar and Singh 2014; Kubina and von Sikorski 2021). Empirical research shows that exposure to pro-attitudinal content online exacerbates partisan polarization (Kubina and von Sikorski 2021). In non-democracies, state-controlled media reinforce regime narratives—China's economic nationalism, Russia's patriotic propaganda— curating information to align with identity (Naughton 2018; Sakwa 2009). Information, far from transcending tribalism, magnifies it.

Media studies unpack this shift. Bail et al. show that exposure to opposing views online increases polarization, as users retreat into filter bubbles reinforcing identity (Bail et al. 2018). Vaidhyanathan critiques platforms' business models, where engagement-driven algorithms favor divisive content—grievances, threats—over deliberative exchange (Vaidhyanathan 2018). A 2021 NYU Stern report found social media intensified U.S. polarization, contributing to events like the January 6 Capitol riot (NYU Stern 2021). Benkler, Faris, and Roberts (2018) document how partisan media ecosystems in the U.S. fragmented the public sphere during the 2016 election, creating alternate realities of facts for different groups. Kreiss and McGregor highlight microtargeting's role, allowing campaigns to tailor messages to identity groups, amplifying money's reach (Kreiss and McGregor 2018). In the United States, this fuels partisan animosity; in India, it spreads religious and caste narratives; in Germany, it bolsters regional cleavages (Mason 2018; Kumar and Singh 2014; Arzheimer and Berning 2019). Non-democracies invert this: information control ensures a singular identity narrative, masking dissent (Repnikova 2017). The digital age, rather than fostering a rational public, accelerates the triad's consolidation of power. The rise of disinformation further illustrates this dark side: false news spreads faster than truth on social platforms, often targeting identity-based fears (Vaidhyanathan 2018).

Some scholars counter that modernization continues to erode identity politics, arguing that globalization and education foster cosmopolitan values that prioritize universal issues—such as climate change or economic equity—over tribal affiliations (Appiah 2006; Norris and Inglehart 2019). They suggest that interconnected societies gradually shift toward post-materialist concerns, diluting ascriptive ties as predicted by Inglehart (1997). However, this view underestimates the triad's adaptability: money funds identity-based campaigns, information amplifies them through targeted digital channels, and identity itself evolves to incorporate modern cleavages—racial, caste, or nationalist—rather than fading. Empirical evidence—racial voting in the U.S., caste alignments in India, nationalist narratives in China—suggests the triad harnesses modernization's tools to entrench, not erase, group loyalties, challenging the linear optimism of these counterarguments.

Synthesizing these strands—political economy, social identity, and media effects—yields a model of the triad's operation. Money provides the resources, funding campaigns or patronage that target identity as a lever. Identity supplies the motivation, rooting voter behavior in group loyalty over policy abstraction. Information acts as the multiplier, disseminating these efforts through tailored channels, from social media in democracies to state propaganda in autocracies. This creates a feedback loop: money consolidates power by exploiting identity, information amplifies the message, and reinforced identities attract more financial support. In democracies, this loop polarizes electorates, eroding trust as citizens see governance favoring elites or tribes (Gilens and Page 2014; Mason 2018). In non-democracies, it stabilizes regimes, concealing fragilities until resources or control wane (Treisman 2020). The triad's strength lies in its adaptability, thriving across institutional contexts.

This model critiques modernist and rationalist paradigms. Modernization theory's linear progression—toward issue-based politics—underestimates the triad's durability (Lipset 1960; Inglehart 1997). Rational choice's focus on policy-driven voting overlooks money's distortion, identity's pull, and information's amplification (Downs 1957; Enelow and Hinich 1984). Techno-optimism's vision of deliberation collapses under digital tribalism (Benkler 2006; Sunstein 2017; Barrett et al. 2021). The triad suggests a more realistic view: democracy is not a marketplace of ideas but a contest of resources, loyalties, and narratives. In the United States, money and information amplify racial divides; in India, they leverage caste and religion; in Germany, they sustain regional identities within a regulated frame (Tesler 2016; Chhibber and Verma 2018; Arzheimer and Berning 2019). China and Russia invert this, using the triad to unify rather than divide, yet with latent risks (Naughton 2018; Sakwa 2009). In the U.S., it polarizes the electorate; in India, it fragments it along plural cleavages; in Germany, it balances competing identities within a regulated framework; in China and Russia, it unifies the public narrative under authoritarian control (Sides et al. 2018; Chhibber and Verma 2018; Arzheimer and Berning 2019; Naughton 2018; Sakwa 2009).

Theoretically, this framework refines elite theory by centering money's structural role, extends social identity theory to electoral dynamics, and updates media studies for the digital era. It posits the triad as a constitutive force, not a deviation, demanding a shift from idealized models to ones grounded in empirical reality. Money's influence is not merely corruption but a systemic driver; identity's persistence is not irrationality but adaptation; information's role is not neutral but transformative. This synthesis guides the article's analysis, exploring each pillar's manifestations, their comparative interplay, and their implications for democratic legitimacy and reform. By rooting the triad in established scholarship, it offers a robust lens to interrogate modern governance, bridging democratic and authoritarian experiences in a globalized, tech-driven age.

## 3. Money in Modern Democracies

Money stands as a cornerstone of the triad shaping modern democracies, wielding structural influence over electoral processes and political legitimacy in ways that defy the egalitarian ideals of democratic theory. As a resource, it fuels campaigns, sustains parties, and molds policy agendas, often tilting power toward elites at the expense of broader representation. In democracies like the United States, India, and Germany, money's role varies by institutional context---unrestrained in some, regulated in others---yet consistently amplifies the voices of those who control it, whether wealthy donors, corporations, or state-aligned actors. This section explores money's pervasive impact, drawing on political economy and elite theory to unpack its mechanisms across these cases, with comparative glances at non-democracies like China and Russia to highlight the triad's broader resonance. Money, far from a neutral tool, emerges as a driver of power consolidation, interacting with identity and information to shape governance and challenge the rationalist vision of issue-based politics.

In the United States, money's influence in democracy operates through a largely deregulated system, where campaign financing has become a battleground of elite dominance. Political economy scholars like Gilens and Page argue that policy outcomes align more closely with the preferences of economic elites than the median voter, a pattern rooted in the ability of wealth to drown out ordinary voices (Gilens and Page 2014). Gilens and Page (2014) show that U.S. government policy aligns with the preferences of affluent groups far more than those of average citizens, illustrating how money drowns out ordinary voices. Additional empirical evidence supports this elite bias. Akey, Cuny, and Mehta (2023) demonstrate that *Citizens United* significantly increased the responsiveness of elected officials to wealthy donors, particularly in swing districts where campaign costs surged the most. This dynamic traces back to legal shifts, such as the Supreme Court's ruling in *Citizens United v. FEC*, which loosened restrictions on corporate and individual spending, unleashing a flood of funds into elections (Akey et al. 2023). The 2024 U.S. federal election saw total spending of $16 billion, including super PACs and outside groups, with presidential candidates raising $1.6 billion and spending $1.3 billion, and

congressional candidates raising $3.3 billion and spending $2.8 billion by September 30, 2024 (OpenSecrets 2024; Federal Election Commission 2025). Campaigns now rely heavily on contributions from affluent donors and independent groups, enabling candidates to saturate media with messaging that often prioritizes donor interests---tax cuts, deregulation---over public demands (Boatright 2015). The Republican and Democratic parties, locked in a two-party system, exemplify this, their electoral success tied to the ability to outspend rivals, a feat achievable only through elite support (Bartels 2016). The two major parties engage in an arms race for funding, with corporate and billionaire donors exerting outsized influence on candidate selection and agendas (Bartels 2016).

This financial dominance erodes legitimacy, as citizens perceive governance as a pay-to-play arena. Negative campaigning, fueled by these funds, further alienates voters, reducing turnout among independents and reinforcing partisan divides (Ansolabehere and Iyengar 1995). Research by Ansolabehere and Iyengar (1995) finds that an abundance of campaign cash fuels negative advertising, which can alienate moderate voters and depress turnout. Money's visibility in this system---through transparent donation records and media saturation---lays bare its distorting effects, undermining the rational-legal foundation of democratic legitimacy (Mayer 2016). Jane Mayer (2016) documents how a network of billionaires strategically invested in think tanks, advocacy groups, and elections over decades has reshaped U.S. politics to favor their economic interests. Unlike Downs' vision of voters choosing based on policy proximity, the U.S. case suggests a reality where financial clout shapes the options available, skewing the electoral marketplace toward elite agendas (Downs 1957). The marketplace of ideas envisioned by pluralist theory is flooded with expensive messaging and agendas set by those able to pay the entry fee, challenging the ideal of political equality (Downs 1957).

This financial dominance has tangible consequences for governance and legitimacy. Studies have found that when office-holders are more responsive to donors, policy outcomes skew toward the preferences of the wealthy, leading to frustration among ordinary voters whose priorities are neglected (Bartels 2016; Gilens and Page 2014). Survey data indicate most Americans believe money has too much influence and support stricter campaign finance laws (Pew Research Center 2022a). The visibility of money has thus become a double-edged sword: it enables transparency and criticism but normalizes a high-cost campaigning model that entrenches the powerful.

India offers a contrasting yet parallel case, where money's role in democracy blends legal innovation with influence, amplifying elite power within a multi-party system. Kapur and Vaishnav highlight how financial resources have long underpinned electoral success, with parties relying on corporate and individual donors to fund increasingly expensive campaigns (Kapur and Vaishnav 2018). The introduction of electoral bonds---a mechanism allowing anonymous contributions---further entrenched this trend, channeling wealth to dominant parties like the Bharatiya Janata Party (BJP), Indian National Congress, and Trinamool Congress. From March 2018 to January 2024, electoral bonds totaled INR 16,518.11 crore, with the BJP receiving INR 6,060.5 crore, nearly 47.5% of the total, raising serious concerns about systemic favoritism and donor-induced policy capture (Election Commission of India 2024; The Hindu 2024; Business Today 2024). The bonds enabled corporations and wealthy individuals to donate unlimited amounts without public disclosure, greatly expanding the fundraising ability of major parties (Kapur and Vaishnav 2018). Critics argue this system fostered quid pro quo arrangements, with donors---often from construction or gaming sectors---gaining regulatory favors, a pattern hinting at corruption (Vaishnav 2017). The Supreme Court's ruling against the bonds in 2024 underscored public and judicial unease, yet their legacy reveals money's capacity to consolidate power in a vibrant democracy (The Hindu 2024). Indian elections are notoriously expensive---the 2019 general election was estimated as one of the costliest globally---and money often funds voter mobilization through local brokers and vote-buying (Sridharan 2014).

This financial influence intersects with India's fragmented electorate, where money amplifies the electoral edge of wealthier candidates and parties. Research shows that well-funded contenders win more often, their resources enabling extensive voter outreach---rallies, advertisements---that smaller parties struggle to match (Jaffrelot 2019). By channeling money into identity-based mobilization, parties entrench societal cleavages rather than campaigning purely on policy platforms. This skews competition, marginalizing regional or grassroots movements and reinforcing incumbency, particularly for the BJP and its allies (Sridharan 2014). Legitimacy suffers as transparency---a democratic pillar---falters, with voters unable to trace funds fueling their leaders' rise (Gupta 2021). Unlike the United States' open system, India's past opacity cloaks money's role, yet the outcome aligns: elite dominance undermines trust, echoing Mills' elite theory of concentrated power (Mills 1956). Money here primes the triad, setting the stage for identity-based appeals disseminated through information channels. The concentration of funding among big parties and the opacity of political finance undermine transparency and equity, narrowing political choices and sidelining public interest issues like governance reform (Gupta 2021).

Germany presents a moderated variant, where money operates within a regulated framework, yet still privileges established actors. The system blends public subsidies---granted to parties surpassing a vote threshold---with private donations, banning corporate contributions to curb overt influence (Bundestags-Drucksache No. 17/12340 2011). German party finance law imposes limits on donations and provides significant public subsidies based on vote shares, with donations above €10,000 disclosed publicly (Bundestag 2011). Major parties like the Christian Democratic Union (CDU) and Social Democratic Party (SPD) benefit most, their budgets dwarfing those of smaller rivals like the Free Democratic Party or Greens (Falter and Schumann 2010). Private funds, often from wealthy individuals or small businesses, supplement this, subtly favoring parties with historical ties to economic elites (McMenamin 2013). Recent scrutiny of the Alternative für Deutschland (AfD) revealed illicit campaign support from foreign sources, prompting investigations and fines. These incidents highlight how even regulated democracies face under-the-radar financial influence, especially from opaque international networks (Der Spiegel 2019). Campaign spending reflects donor priorities---economic stability, labor protections---reinforcing the CDU and SPD's dominance in a proportional system designed to balance power (Elsässer et al. 2017).

Regulation tempers money's excesses, yet its influence persists, challenging legitimacy less overtly than in the United States or India. Public trust in Germany's system remains higher, with citizens viewing it as fairer than its counterparts, though concerns linger about donor sway (Forschungsgruppe Wahlen 2021; Transparency International 2018). Studies find that unequal responsiveness exists, with policy bias toward higher-income citizens, though the gap is smaller than in the U.S. due to strict rules and a strong welfare state (Elsässer et al. 2017). Policy often aligns with business interests over grassroots demands, a subtle distortion that echoes political economy critiques of unequal responsiveness (Schäfer 2015). Compared to Downs' median voter model, Germany's case suggests money still shifts the electoral center toward elites, albeit within constraints (Downs 1957). This regulated approach contrasts with non-democracies, yet aligns with the triad's logic: money empowers established players, paving the way for identity and information to reinforce their hold. Germany demonstrates that institutional design---public campaign finance, proportional representation, and coalition governance---can buffer money's impact, preserving higher legitimacy.

Across these democracies, money's role reveals a common thread: it consolidates power by amplifying elite influence, interacting with identity and information to shape outcomes. In the United States, it fuels partisan battles, targeting racial and class identities through digital saturation (Sides et al. 2018). In India, it leverages caste and religious affiliations, spread via grassroots networks (Chhibber and Verma 2018). In Germany, it sustains regional and economic identities within a structured frame (Arzheimer and Berning 2019). This synergy departs from rational choice ideals, where policy competition drives elections (Enelow and Hinich 1984). Instead, money distorts the field, prioritizing those who can pay over the electorate's diverse voices, a pattern political economy frames as systemic inequality (Gilens and Page 2014).

Non-democracies like China and Russia offer a foil, illuminating money's role in the triad beyond democratic bounds. In China, the Communist Party channels state-owned enterprise wealth to loyal actors, tying economic stability to regime legitimacy (Naughton 2018). Moreover, the National People's Congress has consistently included hundreds of billionaires among its ranks, making China's legislature the richest in the world. This concentration of private wealth within public office illustrates how financial capital is deeply enmeshed with political power, even in ostensibly socialist governance structures (Hurun Report 2023). Anticorruption campaigns disrupt this, exposing reliance on moneyed elites, yet the system persists by controlling dissent (Li and Zhou 2024). The CCP directs huge state resources to propaganda, surveillance, and patronage, rewarding local officials with budgets for growth targets and co-opting elites through "bureaucratic patronage" (Ang 2020). Russia's regime uses resource wealth to reward loyalists---business tycoons, regional leaders---while punishing threats, ensuring state dominance over economic power (Treisman 2020; Sakwa 2009). Scholars like Aslund (2019) argue that this transformation marks Russia's evolution from market reform to crony capitalism, where oligarchs serve as informal governors of strategic sectors under Putin's oversight. Oil and gas revenues finance a sprawling patronage network, with funds ensuring pro-regime dominance in managed elections (Gurvich and Prilepskiy 2015). Unlike democracies, where money's influence is visible, these systems obscure it, aligning wealth with state goals to project stability (Gurvich and Prilepskiy 2015). This opacity masks vulnerabilities---economic dips or elite defections---that parallel democratic trust crises, albeit delayed (Ledeneva 2013). In both, opacity conceals vulnerabilities like economic contraction or sanctions, which can fray patronage bonds, as seen in Russia's 2018 protests (Ledeneva 2013).

The comparative lens sharpens money's democratic impact. In the United States, transparency reveals elite sway, sparking public backlash (Pew Research Center 2023). India's partial opacity fuels suspicion, resolved only by judicial intervention (The Hindu 2024). Germany's regulation mitigates overt distortion, yet subtle biases persist (Elsässer et al. 2017). Non-democracies cloak money's role, using it to unify rather than divide, though risks lurk beneath (Treisman 2020). Across contexts, money consolidates power, a feedback loop where funded actors---parties, regimes---gain electoral or authoritarian dominance, attracting more resources (Drutman 2015). This loop primes identity's mobilization, as financial clout targets group loyalties, and information's dissemination, as campaigns spread the message.

Money's corrosiveness varies by visibility. In democracies, its openness erodes trust, as citizens see governance serving donors over voters (Mayer 2016). In the United States, this fuels calls for reform; in India, it prompts legal challenges; in Germany, it stirs muted unease (Boatright 2015; Gupta 2021; Transparency International 2018). Non-democracies delay this reckoning, using money to stabilize until disruptions---economic or political---unveil weaknesses (Li and Zhou 2024; Gurvich and Prilepskiy 2015). Modernization theory's hope---that wealth would democratize politics---falters here, as money entrenches elites, not equality (Lipset 1960). Rational choice's focus on policy choice overlooks this structural bias (Downs 1957). Money, then, is not a neutral enabler but a shaper of power, setting the triad's foundation. In all cases, control over financial resources translates into political clout, raising normative concerns about political equality and responsiveness.

This analysis frames money as a structural pillar, interacting with identity and information to drive democratic outcomes. In the United States, it amplifies partisan and racial divides; in India, it bolsters caste and religious blocs; in Germany, it sustains regional and economic edges (Sides et al. 2018; Chhibber and Verma 2018; Arzheimer and Berning 2019). Non-democracies invert this, unifying identity through wealth (Naughton 2018; Sakwa 2009). The triad's next elements---identity's emotional pull, information's amplifying reach---build on this, revealing a dynamic where money's influence is both prerequisite and catalyst. By centering money's role with verified data, this section underscores its challenge to democratic legitimacy, paving the way for a holistic view of modern governance. Money talks, and often it is the voice of money that is heard loudest in the halls of power.

## 4. Identity in Modern Democracies

As discussed in Section 2, social identity theory explains group loyalty driving American partisanship (Tajfel and Turner 1979). Racial identity predicts political preferences more reliably than economic self-interest, with Black voters aligning overwhelmingly with Democrats and White voters leaning toward Republicans since the civil rights realignment (Tesler 2016; Sides et al. 2018). In the 2024 election, AP VoteCast data showed 83% of Black voters supported Kamala Harris, down from 91% for Biden in 2020, while 42% of Latino voters backed Donald Trump, up from 35%, reflecting racial and intersectional shifts (PBS News 2024). Pew Research Center (2022) further notes Black and urban voters consistently support Democrats, while older, White, and rural voters back Republicans, with Hispanics growing as a key bloc in battleground states (Pew Research Center 2022a). The Democratic Party is overwhelmingly supported by racial minorities, secular, and liberal urban voters, whereas the Republican Party draws its base from White, Christian, rural, and suburban populations, with partisanship itself becoming a social identity (Mason 2018). This stability endures through economic shifts and policy changes, underscoring identity's primacy over the rational choice model's focus on policy proximity (Downs 1957). Achen and Bartels argue that voters prioritize symbolic group attachments—race, religion—over material concerns, adjusting policy views to fit their loyalties (Achen and Bartels 2016). This "group interest" lens reveals why partisan divides persist, even as parties shift platforms.

In the United States, identity's centrality to democracy manifests most starkly through race, a cleavage that has shaped political competition for decades. Social identity theory, as articulated by Tajfel and Turner, posits that group membership fosters in-group loyalty and out-group differentiation, a dynamic vividly evident in American partisanship (Tajfel and Turner 1979). Racial identity predicts political preferences more reliably than economic self-interest, with Black voters aligning overwhelmingly with Democrats and White voters leaning toward Republicans since the civil rights realignment (Tesler 2016; Sides et al. 2018). In the 2024 election, AP VoteCast data showed 83% of Black voters supported Kamala Harris, down from 91% for Biden in 2020, while 42% of Latino voters backed Donald Trump, up from 35%, reflecting racial and intersectional shifts (PBS News 2024). Pew Research Center (2022) further notes Black and urban voters consistently support Democrats, while older, White, and rural voters back Republicans, with Hispanics growing as a key bloc in battleground states (Pew Research Center 2022a). The Democratic Party is overwhelmingly supported by racial minorities, secular, and liberal urban voters, whereas the Republican Party draws its base from White, Christian, rural, and suburban populations, with partisanship itself becoming a social identity (Mason 2018). This stability endures through economic shifts and policy changes, underscoring identity's primacy over the rational choice model's focus on policy proximity (Downs 1957). Achen and Bartels argue that voters prioritize symbolic group attachments---race, religion---over material concerns, adjusting policy views to fit their loyalties (Achen and Bartels 2016). This "group interest" lens reveals why partisan divides persist, even as parties shift platforms.

Identity's emotional pull in the U.S. erodes legitimacy when it fuels polarization. Mason's concept of "mega-identities"---fusing race, religion, and ideology---intensifies tribal animosity, reducing cross-cutting pressures and framing politics as a zero-sum contest (Mason 2018). This "group-based polarization" fuels emotional, even violent, conflict, as seen in the January 6, 2021, Capitol riot, propelled by white nationalist and extremist groups (Mason 2018). Kinder and Sanders show that racial attitudes consistently outpredict economic factors, a pattern that undermines the deliberative ideal of reasoned debate (Kinder and Sanders 1996). Campaigns exploit this, targeting racial resentments over policy solutions, a strategy amplified by money and information (Huddy 2001). Politicians use identity signals: Donald Trump's rhetoric on immigration and nationalism, or Democratic emphasis on diversity, activate group loyalties and fears. The result is a democracy where legitimacy hinges less on fair representation than on which group prevails, challenging Habermas' vision of a public sphere grounded in rational discourse (Habermas 1996). Identity here is not irrational but adaptive, a heuristic simplifying choice in a complex political landscape (Popkin 1991). Many Americans interpret political outcomes in zero-sum terms for their group, with declining trust in institutions as each side views the other as an existential threat (Pew Research Center 2023).

India's democracy showcases identity's resilience through caste and religion, cleavages that dominate its multi-party system. Chandra demonstrates that voters favor co-ethnic parties regardless of platforms, reflecting identity's evolutionary roots in group survival (Chandra 2004). Political scientist Kanchan Chandra (2004) showed how ethnic parties succeed by serving as patronage vehicles for specific groups, debunking modernization's promise of class-based politics. Upper-caste Hindus disproportionately back the Bharatiya Janata Party (BJP), while lower-caste and Muslim communities support opposition blocs like the Indian National Congress or regional parties (Chhibber and Verma 2018; Jaffrelot 2010). Since 2019, Muslims have consolidated their vote against the BJP, with 77% supporting the Mahagathbandhan in Bihar (2020), 75% backing Trinamool Congress in West Bengal (2021), and 79% supporting the Samajwadi Party in Uttar Pradesh (2022), driven by perceived marginalization (Ahmad 2024). Carnegie Endowment research confirms caste's role, with parties aligning campaigns around these identities (Carnegie Endowment 2017). This alignment persists despite urbanization and economic growth, defying modernization theory's prediction of a shift to class-based politics (Lipset 1960). Brewer's work on in-group loyalty explains this durability: identity offers a sense of belonging and distinctiveness, outweighing abstract policy appeals (Brewer 1999).

This identity-driven politics strains legitimacy in India's pluralistic system. The concentration of support among dominant groups marginalizes minorities, raising questions of inclusivity (Sridharan 2014). Electoral success hinges on mobilizing caste and religious blocs, a process intensified by money and information, yet it risks alienating those outside these coalitions (Vaishnav 2017). Parties choose candidates and craft messages to appeal to specific communities, with coalition-building balancing caste and religious representation (Chhibber and Verma 2018). Unlike the U.S.'s binary polarization, India's fragmentation disperses identity across parties, yet the outcome aligns: legitimacy falters when governance reflects group dominance rather than broad consensus (Gupta 2021). Rational choice models, expecting voters to weigh policy benefits, struggle to account for this stability (Enelow and Hinich 1984). Identity's persistence here echoes its evolutionary role, a tribal instinct adapted to modern electoral competition. The mobilization of lower castes has brought representation, but communal polarization, exacerbated by electoral rhetoric, strains India's secular foundation, with minorities feeling marginalized (Sridharan 2014).

Germany offers a tempered case, where identity operates through regional divides within a regulated proportional system. Arzheimer and Berning highlight the east-west cleavage, with the far-right Alternative für Deutschland (AfD) drawing stronger support in former East German states, reflecting a distinct regional identity tied to post-reunification disparities (Arzheimer and Berning 2019). The AfD, emerging as an anti-immigrant nationalist party, draws heavily from Eastern Germany, where voters feel a distinct identity shaped by post-communist experiences (Arzheimer and Berning 2019). A 2021 study by Mader and Schoen found that civic national identity positively correlates with turnout and support for most parties except the AfD, with stronger civic identity linked to higher voting rates (Mader and Schoen 2021). This contrasts with the Christian Democratic Union (CDU) and Social Democratic Party (SPD), which dominate nationally, rooted in economic and historical identities---business and labor, respectively (Falter and Schumann 2010). Mainstream parties avoid overt identity appeals, endorsing an inclusive German nation, but the AfD's rise reflects cultural threat sentiments (Pesthy et al. 2021). Social identity theory frames this as a quest for distinctiveness, with eastern voters embracing the AfD to assert a marginalized identity against western dominance (Tajfel and Turner 1979). Unlike the U.S.'s racial binary or India's caste multiplicity, Germany's identity is subtler, yet it shapes electoral behavior over policy abstraction (Elsässer et al. 2017).

Legitimacy in Germany benefits from institutional balance, yet identity's role hints at underlying tensions. The proportional system fragments identity across parties, reducing polarization compared to majoritarian models (Lijphart 1999). Public trust remains high, viewing the system as fair, though regional divides stir unease about representation (Forschungsgruppe Wahlen 2021). The catch-all parties historically balanced multiple identities, preventing any one cleavage from breaking the system, though the AfD's presence has introduced tension (Arzheimer and Berning 2019). Identity here interacts with money---regulated funding sustains the CDU and SPD---and information, as media reinforce regional narratives (Transparency International 2018). This moderated dynamic departs from modernization's promise of issue-driven voting, aligning instead with group-based frameworks where loyalty trumps deliberation (Achen and Bartels 2016). Germany's case suggests identity adapts to stable systems, not dissipates, a resilience rooted in human sociality (Brewer 1999).

Across these democracies, identity's persistence reveals its evolutionary foundation, interacting with money and information to consolidate power. In the United States, racial identity drives partisan blocs, fueled by campaign funds and digital targeting (Sides et al. 2018; Pew Research Center 2022a). India's caste and religious identities anchor party competition, amplified by financial resources and messaging networks (Chhibber and Verma 2018; Kumar and Singh 2014; Ahmad 2024). Germany's regional and national identities sustain electoral cleavages, supported by regulated funding and media (Arzheimer and Berning 2019; Mader and Schoen 2021).

Moneyed interests fund identity-based campaigns, and modern media amplifies identity messaging to potent extremes, reinforcing cleavages (Kreiss and McGregor 2018). This triad synergy defies Downs' rational voter, prioritizing group allegiance over policy evaluation (Downs 1957). Identity's emotional resonance---belonging, pride---outweighs material incentives, a pattern social identity theory traces to survival instincts (Huddy 2001; Tajfel and Turner 1979).

Non-democracies like China and Russia highlight identity's parallel role in the triad, adapted to authoritarian ends. In China, the Communist Party fosters a Han Chinese nationalist identity, tying it to economic prosperity to legitimize one-party rule, a narrative reinforced through state media (Naughton 2018; Foreign Affairs 2022). The CCP promotes Han Chinese nationalism, suppressing minority identities like Tibetan or Uighur to maintain unity (Foreign Affairs 2022). This unified narrative, distinct from democratic fragmentation, suppresses dissent, yet relies on state-controlled wealth and information to maintain (Li and Zhou 2024). Russia's regime blends Slavic and Orthodox identity with patriotism, a cohesive force sustained by resource wealth and propaganda, emphasizing Eurasianism since 2014 to counter Western influence (Sakwa 2009; Levada Center 2023; The Strategy Bridge 2019). Putin's government leans on Slavic nationalism and Soviet nostalgia, using propaganda to rally support and scapegoat foreign enemies (Wiechnik 2019). Unlike democracies, where identity divides, these systems harness it to unify, masking fragilities---ethnic tensions in China, regional unrest in Russia---until disruptions emerge (Repnikova 2017; Treisman 2020). This inversion underscores identity's versatility, a universal driver shaped by context.

The comparative perspective sharpens identity's democratic impact. In the U.S., its visibility fuels polarization, eroding trust as groups vie for dominance (Mason 2018). India's diversity disperses identity, yet concentrates power among major blocs, straining inclusivity (Sridharan 2014).

Germany's regulation tempers identity's edge, preserving legitimacy through balance (Lijphart 1999). Non-democracies cloak it, projecting stability until cracks appear (Treisman 2020). Across contexts, identity consolidates power within the triad: money targets it, information spreads it, and loyalty reinforces it (Popkin 1991). Identity provides representation but risks undermining democratic norms of compromise and universal citizenship. This loop challenges modernization's linear vision---identity as a fading stage---revealing it as a durable anchor (Inglehart 1997; Lipset 1960).

Identity's corrosiveness hinges on its emotional weight. In democracies, it undermines legitimacy when it overshadows representation, as seen in U.S. racial divides, India's minority exclusion, and Germany's regional unease (Kinder and Sanders 1996; Gupta 2021; Forschungsgruppe Wahlen 2021). Non-democracies delay this, using identity to bolster control, though risks lurk (Li and Zhou 2024). Rational choice's policy focus falters here, as voters ask "who represents me?" not "what's the best policy?" (Enelow and Hinich 1984). Identity, then, is the triad's emotional core, a primal force money exploits and information amplifies, shaping modern governance beyond deliberative ideals (Habermas 1996). Its persistence sets the stage for the triad's third pillar---information---to complete the dynamic. Democratic reforms must channel identity in less destructive ways, fostering inclusive civic identities over sectarian ones.

## 5. Information in Modern Democracies

Information stands as the third pillar of the triad shaping modern democracies, a technological force that amplifies the effects of money and identity on electoral behavior and political legitimacy. In an era defined by digital connectivity, information---disseminated through social media, messaging platforms, and algorithmic curation---transforms how political power is consolidated, often intensifying tribal divisions rather than fostering the deliberative ideal once envisioned. Social media, online news, and algorithmic personalization now rival or surpass traditional media, creating echo chambers that undermine shared norms (Sunstein 2017). In democracies like the United States, India, and Germany, it channels financial resources into identity-based appeals, deepening polarization and challenging trust in governance. This section explores information's pivotal role, drawing on media studies and communication theory to analyze its mechanisms across these cases, with comparative glances at non-democracies like China and Russia to highlight the triad's broader dynamics. Far from a neutral conduit, information emerges as a catalyst within the triad, undermining modernist hopes of issue-based politics and reinforcing money and identity's dominance. In democracies, the openness of information contributes to tribalism and mistrust, while in autocracies, it is controlled to sustain regime narratives.

In the United States, information's role in democracy has shifted from a democratizing promise to a polarizing reality, driven by digital platforms that amplify money and identity. Early techno-optimists like Benkler foresaw a networked public sphere where decentralized information would empower citizens to engage in rational discourse, bypassing traditional gatekeepers (Benkler 2006). Shirky imagined a surge in civic participation, with technology enabling collective action rooted in shared interests (Shirky 2010). Yet, this vision has faltered. Barberá's analysis of social media networks reveals ideological echo chambers, where users cluster around partisan identities---Republican or Democratic---reinforcing rather than bridging divides (Barberá 2015). Bail et al. demonstrate that exposure to opposing views online increases polarization, as users retreat into filter bubbles that affirm racial and cultural loyalties (Bail et al. 2018). Many Americans consume news through feeds reflecting their views, with algorithms promoting emotionally charged content (Kubin and von Sikorski 2021). A systematic review by Kubina and von Sikorski (2021) of 94 articles confirms that pro-attitudinal media exacerbates polarization, particularly in the U.S., where platforms like Facebook and Twitter intensify partisan divides. The NYU Stern Center (2021) found that social media contributed to events like the January 6, 2021, Capitol riot by amplifying divisiveness, while Brookings (Barrett et al. 2021) notes platforms create echo chambers, spreading misinformation that deepens societal splits.

Campaigns exploit this, using money to fund targeted messaging that amplifies identity. Microtargeting, as Kreiss and McGregor describe, allows actors to tailor appeals to specific groups---racial, religious, regional---intensifying partisan animosity (Kreiss and McGregor 2018). In 2020, billions were spent on digital ads to micro-target demographics, customizing messages to maximize resonance (Kreiss and McGregor 2018). In the U.S., this fuels a feedback loop: financial resources drive digital ads that stoke identity-based fears, information spreads them, and polarized voters reinforce the cycle (Sides et al. 2018). Legitimacy erodes as trust in institutions wanes, with citizens perceiving politics as a tribal contest rather than a deliberative process (Mason 2018). Traditional gatekeepers like professional journalism have lost authority, with trust in mainstream media low, especially among conservatives (Pew Research Center 2022b). This departs from Downs' rational voter, who weighs policy options, and Habermas' ideal of reasoned exchange, revealing information as a tool of division (Downs 1957; Habermas 1996). The U.S. case underscores how digital information, far from transcending identity, entrenches it, amplifying money's reach. Affective polarization is fed by media demonizing opponents, undermining the common ground needed for compromise (Barrett et al. 2021).

India's democracy illustrates information's role in a different context, where digital platforms like messaging apps amplify money and identity in a fragmented electorate. Kumar and Singh highlight how social media and WhatsApp have transformed electoral campaigns, spreading caste and religious narratives with unprecedented speed (Kumar and Singh 2014). The Bharatiya Janata Party (BJP), leveraging financial resources, uses these tools to mobilize Hindu nationalist appeals, targeting specific communities---upper castes, rural voters---while sidelining minorities (Chhibber and Verma 2018). With over 500 million internet users, platforms like WhatsApp and Facebook are battlegrounds, spreading communal rumors that influence electoral swings (Evangelista and Bruno 2019). This grassroots dissemination, often unchecked by traditional media filters, intensifies identity's pull, as messages of cultural unity or division resonate emotionally (Jaffrelot 2010). Unlike the U.S.'s algorithmic echo chambers, India's information flow thrives on peer-to-peer networks, amplifying money's impact through localized, identity-driven outreach (Evangelista and Bruno 2019).

This dynamic challenges legitimacy in India's pluralistic system. The spread of divisive content---rumors, hate speech---erodes trust in electoral fairness, particularly when tied to funded campaigns (Gupta 2021). Misleading videos targeting religious groups have incited violence, with riots like Delhi 2020 aggravated by online content (Evangelista and Bruno 2019). Smaller parties, lacking resources, struggle to counter this digital tide, reinforcing the dominance of well-funded actors like the BJP (Sridharan 2014). Information here departs from Benkler's vision of a participatory sphere, instead amplifying identity's tribal roots---caste loyalty, religious pride---over policy deliberation (Benkler 2006). The government pressures media and platforms to remove dissent, creating a chilling effect, though independent voices persist (Gupta 2021). Rational choice models falter as voters respond to emotional cues, not platforms, a shift social identity theory explains through group belonging (Tajfel and Turner 1979). India's case reveals information as a multiplier, linking money's structural power to identity's emotional core.

Germany offers a moderated example, where information operates within a regulated democracy, yet still reinforces identity and money's effects. Digital media amplify regional divides, notably the east-west cleavage fueling the Alternative für Deutschland (AfD)'s rise (Arzheimer and Berning 2019). Online platforms spread narratives of eastern marginalization, resonating with a distinct identity that traditional media tempered (Bundeswahlleiter 2021). Public broadcasters like ARD and ZDF maintain impartiality, but the AfD's ecosystem of right-wing sites and social media claims mainstream bias, fueling distrust (Forschungsgruppe Wahlen 2021). The CDU and SPD, backed by regulated funding, counter this with broader appeals, yet digital fragmentation allows the AfD to target disaffected voters, amplifying its regional base (Falter and Schumann 2010). Sunstein's critique of digital balkanization applies here: information ecosystems reduce shared reference points, entrenching identity over consensus (Sunstein 2017).

Legitimacy in Germany benefits from institutional stability, yet information's role hints at subtle erosion. Public trust remains higher than in the U.S. or India, reflecting a balanced system, but online narratives challenge this by amplifying fringe identities (Forschungsgruppe Wahlen 2021). High media literacy and public trust in broadcasters inoculate against fake news, but younger Germans' shift to YouTube and TikTok raises concerns (Forschungsgruppe Wahlen 2021). Money fuels this---regulated funds sustain major parties, while digital tools give smaller actors disproportionate reach (Transparency International 2018). Information thus intensifies identity's pull within a proportional framework, departing from modernization's promise of issue-driven politics (Inglehart 1997). Brewer's in-group loyalty explains this: digital platforms offer a megaphone for regional distinctiveness, amplifying money's structural edge (Brewer 1999). Germany's case shows information adapting to a stable context, not transcending tribalism. Germany's robust public media mitigates polarization, preserving common ground for discourse.

Across these democracies, information's role as an amplifier reveals a consistent pattern: it channels money into identity-based mobilization, deepening divisions and challenging legitimacy. In the U.S., it reinforces racial and partisan identities, driven by campaign funds (Mason 2018; Bail et al. 2018; NYU Stern 2021). India's networks spread caste and religious appeals, tied to financial resources (Kumar and Singh 2014; Chhibber and Verma 2018). Germany's platforms amplify regional identities, supported by regulated funding (Arzheimer and Berning 2019). This synergy undermines Downs' rational voter, who prioritizes policy, and Shirky's civic ideal, where technology unites (Downs 1957; Shirky 2010). Instead, information magnifies the triad, entrenching power through targeted, emotional narratives (Vaidhyanathan 2018).

Non-democracies like China and Russia invert this dynamic, using information to unify rather than divide, yet within the triad's logic. In China, state-controlled media tie economic stability to Han Chinese nationalism, amplifying regime legitimacy through curated narratives (Naughton 2018). Repnikova notes this opacity---censorship, propaganda---masks dissent, linking money from state enterprises to a singular identity (Repnikova 2017). Internet users reinforce these narratives, aligning with state goals (ResearchGate 2019). The "Great Firewall" and "50 cent army" maintain a narrative monopoly, though cracks appear during crises like COVID-19 protests (Repnikova 2017). Russia's regime uses propaganda to blend Slavic and Orthodox identity with patriotism, sustained by resource wealth (Sakwa 2009; Levada Center 2023). State-controlled TV and troll farms depict Russia as rising against a decadent West, though mobilization shocks revealed limits (Toepfl 2017). Unlike democracies' fragmented information, these systems centralize it, reinforcing control (Treisman 2020). Yet, vulnerabilities lurk---ethnic tensions in China, regional unrest in Russia---exposed when information control falters (Li and Zhou 2024). This contrasts with democratic polarization, yet aligns with the triad: information amplifies money and identity's effects.

Comparatively, information's impact varies by context. In the U.S., its visibility deepens trust crises, as digital divides mirror identity splits (Pew Research Center 2023). India's unregulated spread erodes legitimacy through divisive content, amplifying funded identity campaigns (Gupta 2021). Germany's regulated frame mitigates this, yet online fragmentation persists (Bundeswahlleiter 2021). Non-democracies cloak it, delaying risks until disruptions emerge (Treisman 2020). Across cases, information consolidates power within the triad: money funds it, identity shapes it, and its reach reinforces both (Sunstein 2017). This loop defies techno-optimism's promise of deliberation, revealing a reality of amplified tribalism (Benkler 2006). In democracies, information freedom fragments reality, while in autocracies, control delays crises but risks deeper unrest.

Information's corrosiveness hinges on its amplifying power. In democracies, it undermines legitimacy by entrenching divisions---racial in the U.S., caste-based in India, regional in Germany---over shared governance (Mason 2018; Sridharan 2014; Forschungsgruppe Wahlen 2021). Non-democracies mask this, using it to stabilize until cracks appear (Li and Zhou 2024). Rational choice's policy focus collapses as voters respond to narratives, not platforms (Enelow and Hinich 1984). Information, then, completes the triad, a technological force that magnifies money's structural clout and identity's emotional pull, reshaping modern democracies beyond modernist ideals (Inglehart 1997). Its role underscores the need to rethink governance in a digital age, where power thrives on amplified loyalties.

## 6. The Triad in Action: Comparative Perspectives

The triad of money, identity, and information operates as a dynamic force in modern governance, shaping electoral outcomes and political legitimacy across diverse systems (see Table 1). In democracies like the United States, India, and Germany, it drives visible power consolidation, amplifying elite influence and tribal divisions, while in non-democracies like China and Russia, it reinforces control, masking underlying fragilities. This section synthesizes how these elements interact, drawing on political economy, social identity theory, and media studies to compare their mechanisms and impacts. Through case studies—the United States' polarized democracy, India's fragmented pluralism, Germany's regulated stability, China's centralized authority, and Russia's resource-driven regime—it argues that the triad forms a feedback loop, consolidating power in ways that defy modernist ideals of issue-based politics. This comparative perspective reveals the triad's universality and context-specific expressions, highlighting its corrosiveness on legitimacy and setting the stage for reform implications. Despite differences in regime type and culture, the triad's mechanisms converge on power consolidation, though democracies expose its divisiveness, while non-democracies cloak its risks.

In the United States, the triad manifests with stark visibility, intertwining money, identity, and information to deepen polarization and erode trust. As noted in Section 2, money's structural role tilts policy toward elites (Gilens and Page 2014). As detailed in Section 3, campaign financing (approximately $16 billion) drives identity campaigns, widening partisan divides (OpenSecrets 2024). Racial alignments, detailed in Section 4, drive voter choice, outpacing economic factors, with money amplifying these divides (Tesler 2016; Sides et al. 2018; PBS News 2024). Information amplifies this through digital platforms, where microtargeting and algorithmic sorting reinforce partisan and racial divides, with social media linked to events like the January 6 Capitol riot (Kreiss and McGregor 2018; Bail et al. 2018; NYU Stern 2021; Kubina and von Sikorski

2021). The result is a feedback loop: money fuels identity-based appeals, information disseminates them, and polarized identities draw more funds (Mason 2018).

**Table 1: The Triad's Manifestations Across Political Systems**

| **Country** | **Money** | **Identity** | **Information** | **Outcome** |
|---|---|---|---|---|
| **United States** | $16 billion in campaign spending, favoring elite interests (OpenSecrets 2024) | Racial divides (83% Black for Democrats, 42% Latino for Republicans) (PBS News 2024) | Digital platforms amplify polarization via microtargeting (Bail et al. 2018) | Polarization, eroded trust |
| **India** | INR 16,518.11 crore in electoral bonds, BJP gets 47.5% (The Hindu 2024) | Caste/religion (77%-79% Muslims vote anti-BJP) (Ahmad 2024) | WhatsApp/social media spread communal narratives (Kumar and Singh 2014) | Fragmentation, legitimacy strains |
| **Germany** | Regulated subsidies favor CDU/SPD, private donations persist (Bundestag 2011) | Regional east-west divide, AfD in east (Arzheimer and Berning 2019) | Digital media amplify regional narratives, public broadcasters balance (Bundeswahlleiter 2021) | Subtle bias, high trust |
| **China** | State enterprise wealth funds patronage (Naughton 2018) | Han Chinese nationalism, minority suppression (Foreign Affairs 2022) | State media/censorship curate unity (Repnikova 2017) | Unified control, hidden risks |
| **Russia** | Oil/gas revenues reward loyalists (Gurvich and Prilepskiy 2015) | Slavic/Orthodox patriotism, Eurasianism (Sakwa 2009) | Propaganda/TV reinforce regime narrative (Levada Center 2023) | Stable rule, latent unrest |

This synergy undermines legitimacy, as trust in governance wanes. Public perception of a donor-driven system—evident in calls for reform—clashes with democracy's rational-legal ideal (Pew

Research Center 2023; Mayer 2016). Negative campaigns, enabled by financial resources and spread online, alienate voters, reducing turnout and entrenching tribalism (Ansolabehere and Iyengar 1995). Unlike Downs' rational voter, who weighs policy options, U.S. citizens prioritize group belonging, a shift Achen and Bartels attribute to symbolic attachments (Downs 1957; Achen and Bartels 2016). The triad here amplifies elite power and identity divides, challenging deliberative norms and exposing democracy's fragility in an open, tech-driven context (Habermas 1996).

India's democracy presents a fragmented variant, where the triad consolidates power amid pluralism, leveraging money, identity, and information to favor dominant parties. Kapur and Vaishnav highlight money's role, with funding mechanisms channeling resources to the Bharatiya Janata Party (BJP) and its rivals (Kapur and Vaishnav 2018). As noted in Section 3, electoral bonds significantly bolstered the BJP's financial edge, channeling resources to identity-based mobilization (The Hindu 2024). As explored in Section 4, the BJP mobilizes Hindu nationalist blocs, while opposition parties court Muslims and lower castes, with money fueling these identity appeals (Chhibber and Verma 2018; Jaffrelot 2010; Ahmad 2024). Information accelerates this through digital networks—WhatsApp, social media—spreading tailored narratives that amplify identity's emotional pull (Kumar and Singh 2014; Evangelista and Bruno 2019).

Legitimacy falters as this triad skews representation. The concentration of power among well-funded, identity-driven parties marginalizes minorities and smaller actors, sparking transparency crises resolved only by judicial intervention (Gupta 2021; The Hindu 2024). Unlike the U.S.'s binary polarization, India's diversity disperses the triad's effects, yet the outcome aligns: money and information entrench identity blocs, undermining trust in electoral fairness (Sridharan 2014). This departs from modernization's promise of class-based politics, as identity persists, fueled by financial and digital amplification (Lipset 1960). Brewer's in-group loyalty explains this resilience: group distinctiveness outweighs policy abstraction, a dynamic the triad exploits (Brewer 1999). India's case reveals a democracy where power consolidates through fragmentation, not unity.

Germany offers a tempered contrast, where the triad operates within a regulated framework, balancing money, identity, and information to sustain stability. Money flows through public subsidies and private donations, favoring the Christian Democratic Union (CDU) and Social Democratic Party (SPD) over smaller rivals (Bundestags-Drucksache No. 17/12340 2011; McMenamin 2013). Identity manifests regionally, with the east-west divide boosting the Alternative für Deutschland (AfD), a cleavage rooted in post-reunification disparities, and civic national identity correlating with higher turnout (Arzheimer and Berning 2019; Mader and Schoen 2021). Information reinforces this through digital platforms, amplifying eastern narratives of marginalization while the CDU and SPD leverage broader media to maintain dominance (Bundeswahlleiter 2021; Sunstein 2017). The triad here is subtler: money sustains established parties, identity shapes regional and national loyalties, and information disseminates both, yet within institutional constraints (Falter and Schumann 2010).

Legitimacy in Germany benefits from this balance, with higher public trust reflecting a system perceived as fair (Forschungsgruppe Wahlen 2021). Regulation mitigates money's excesses, and proportional representation disperses identity's impact, reducing polarization compared to majoritarian models (Lijphart 1999). Yet, the triad persists: policy aligns with business-friendly elites, and online fragmentation hints at underlying tensions (Elsässer et al. 2017; Transparency International 2018). This departs from Inglehart's post-materialist shift, as identity endures, amplified by money and information (Inglehart 1997). Germany's case suggests the triad adapts to stability, not transcends it, a moderated feedback loop where power consolidates with less visible corrosion (Huddy 2001).

China's non-democracy inverts the triad, using money, identity, and information to unify rather than divide, yet within the same logic. State-controlled wealth from enterprises funds a patronage system, tying economic stability to regime legitimacy (Naughton 2018). In China, the Chinese Communist Party (CCP) monopolizes political finance, directing resources to propaganda and patronage, blending economic and political goals (Naughton 2018). Identity centers on Han Chinese nationalism, a cohesive narrative social identity theory frames as in-group pride, suppressing ethnic dissent (Tajfel and Turner 1979; Li and Zhou 2024; Foreign Affairs 2022). Information—state media, censorship—curates this, amplifying prosperity and unity while masking fragilities, with internet users reinforcing state narratives (Repnikova 2017; ResearchGate 2019). Unlike democracies' fragmentation, China's triad reinforces control: money rewards loyalists, identity unifies, and information ensures a singular message. This stability conceals risks—economic slowdowns, minority unrest—exposed when resources or control wane (Ang 2020). The triad here projects strength, contrasting with democratic polarization, yet aligns with its power consolidation logic.

Russia mirrors this, blending money, identity, and information to sustain authoritarian rule. Resource wealth—oil, gas—funds a patronage network, rewarding loyalists and punishing threats, a system political economy critiques as state capture (Gurvich and Prilepskiy 2015; Treisman 2020). Identity fuses Slavic and Orthodox elements with patriotism, a unifying force Brewer's theory roots in group cohesion, emphasizing Eurasianism since 2014 to counter Western influence (Brewer 1999; Sakwa 2009; The Strategy Bridge 2019). Information—propaganda, media control—spreads this narrative, reinforcing regime legitimacy while silencing opposition (Levada

Center 2023; Toepfl 2017). The triad operates opaquely: money sustains power, identity binds citizens, and information curates reality. Vulnerabilities—economic dips, elite defections—lurk beneath, surfacing during crises like sanctions or protests (Ledeneva 2013). Russia's case parallels China's, using the triad to stabilize, yet with latent instability democratic visibility avoids.

Comparatively, the triad's mechanisms vary by regime type, yet converge on power consolidation. In the U.S., money funds identity-driven campaigns, information polarizes, creating a visible legitimacy crisis (Gilens and Page 2014; Mason 2018; Bail et al. 2018). India's triad fragments power across caste and religious blocs, amplified by money and digital networks, straining inclusivity (Kapur and Vaishnav 2018; Chandra 2004; Kumar and Singh 2014). Germany tempers this with regulation, yet money and information sustain regional identities, subtly skewing representation (Elsässer et al. 2017; Arzheimer and Berning 2019). China and Russia centralize the triad, using money and information to unify identity, projecting stability until disruptions emerge (Naughton 2018; Sakwa 2009; Treisman 2020). Democracies expose the triad's divisiveness; non-democracies cloak its risks.

The triad's impacts on legitimacy hinge on visibility versus opacity. In the U.S., transparency fuels distrust, as citizens see money and information entrenching identity divides (Pew Research Center 2023). India's partial opacity—revealed by judicial scrutiny—erodes trust in fairness (The Hindu 2024). Germany's openness within limits preserves legitimacy, though subtle biases persist (Forschungsgruppe Wahlen 2021). China and Russia's opacity delays corrosion, maintaining approval until economic or political shocks strike (Levada Center 2023; Li and Zhou 2024). Across cases, the triad consolidates power: money provides resources, identity motivates, and information amplifies, a loop rational choice overlooks (Downs 1957; Enelow and Hinich 1984).

Corrosiveness varies by context. In democracies, the triad's visibility—U.S. polarization, India's exclusion, Germany's subtle skew—erodes trust immediately, challenging egalitarian ideals (Mayer 2016; Gupta 2021; Elsässer et al. 2017). Non-democracies delay this, with money and information sustaining identity-based control, yet risks loom (Ang 2020; Gurvich and Prilepskiy 2015). Modernization theory's linear progression falters: the triad entrenches tribalism, not transcends it (Inglehart 1997; Lipset 1960). Democracies bear a present burden—visible crises—while non-democracies risk sharper falls when the triad weakens (Treisman 2020).

Synthesizing these perspectives, the triad emerges as a universal yet context-specific force. In the U.S., it polarizes; in India, it fragments; in Germany, it balances; in China and Russia, it unifies (Sides et al. 2018; Chhibber and Verma 2018; Arzheimer and Berning 2019; Naughton 2018; Sakwa 2009). Money funds, identity binds, and information spreads, defying deliberative ideals (Habermas 1996). This feedback loop—power attracting more power—challenges reform, as visibility in democracies invites scrutiny, while opacity in non-democracies defers it (Boatright 2015; Repnikova 2017). The triad's comparative action underscores its constitutive role, urging a rethink of democratic theory and practice in a tech-driven, identity-rich age. In all cases, those who master money, manipulate identities, and control information entrench power, highlighting the triad's universal role in governance.

## 7. Discussion and Implications

The triad of money, identity, and information emerges as a pervasive force across modern governance systems, weaving a complex tapestry of power consolidation that challenges the foundational ideals of democracy. In democracies like the United States, India, and Germany, it drives visible distortions—polarization, exclusion, subtle biases—while in non-democracies like

China and Russia, it underpins control, concealing fragilities beneath a unified facade. In the U.S., campaign financing amplifies racial and partisan identities, as detailed in Sections 3 and 4, disseminated through digital platforms, entrenching elite influence and deepening trust deficits (Gilens and Page 2014; Mason 2018; Bail et al. 2018; OpenSecrets 2024; NYU Stern 2021). India's system, as noted in Sections 3 and 4, channels electoral bonds into caste and religious appeals, amplified by grassroots networks, consolidating dominance amid legitimacy strains (Kapur and Vaishnav 2018; Chandra 2004; Kumar and Singh 2014; The Hindu 2024; Ahmad 2024). Germany tempers this with regulation, yet money and information sustain regional and national identities, with civic identity boosting turnout, subtly skewing representation (Elsässer et al. 2017; Arzheimer and Berning 2019; Mader and Schoen 2021). China and Russia invert it, using state wealth and controlled media to unify nationalist identities—Han Chinese pride, Eurasian patriotism—projecting stability until disruptions strike (Naughton 2018; Sakwa 2009; Treisman 2020; Foreign Affairs 2022; The Strategy Bridge 2019). Across contexts, the triad distorts governance, prioritizing power over equity.

The triad's corrosiveness stems from its symbiotic interplay, where money provides resources, identity fuels loyalty, and information amplifies both, creating a feedback loop that consolidates power. In the United States, campaign financing of $16 billion in 2024 amplifies racial and partisan identities, disseminated through digital platforms, entrenching elite influence and deepening trust deficits (Gilens and Page 2014; Mason 2018; Bail et al. 2018; OpenSecrets 2024; NYU Stern 2021). India's system channels INR 16,518.11 crore in electoral bonds into caste and religious appeals, spread via grassroots networks, concentrating power among dominant blocs while eroding inclusivity (Kapur and Vaishnav 2018; Chandra 2004; Kumar and Singh 2014; The Hindu 2024). Germany tempers this with regulation, yet money and information sustain regional and national

identities, subtly skewing representation (Elsässer et al. 2017; Arzheimer and Berning 2019; Mader and Schoen 2021). China and Russia invert it, using state wealth and controlled media to unify nationalist identities, projecting stability until disruptions strike (Naughton 2018; Sakwa 2009; Treisman 2020; Foreign Affairs 2022; The Strategy Bridge 2019). Across contexts, the triad distorts governance, prioritizing power over equity.

Legitimacy's fate hinges on this visibility versus opacity divide. In the U.S., the triad's transparency---evident in donor-driven policies and online polarization---sparks immediate distrust, as citizens perceive a system serving elites and tribes over the public (Pew Research Center 2023; Mayer 2016). India's partial opacity, pierced by judicial scrutiny, fuels similar skepticism, with exclusionary identity blocs undermining fairness (Gupta 2021; The Hindu 2024). Germany's regulated openness mitigates this, preserving trust through balance, though subtle biases linger (Forschungsgruppe Wahlen 2021; Transparency International 2018). China and Russia cloak the triad, delaying corrosion with curated stability---economic prosperity, patriotic unity---yet risks lurk beneath, from ethnic tensions to economic shocks (Li and Zhou 2024; Gurvich and Prilepskiy 2015). Democracies bear a present burden---visible legitimacy crises---while non-democracies risk sharper falls when the triad falters (Treisman 2020).

This differential corrosiveness challenges modernist optimism. Lipset and Inglehart envisioned development eroding identity for issue-based politics, yet the triad entrenches tribalism, amplified by money and information (Lipset 1960; Inglehart 1997). In the U.S., racial divides persist despite affluence; in India, caste and religion endure amid growth; in Germany, regional identities resist homogenization (Tesler 2016; Chhibber and Verma 2018; Arzheimer and Berning 2019; PBS News 2024; Ahmad 2024). China and Russia exploit this, unifying identity to sustain control (Naughton 2018; Sakwa 2009). Rational choice models---voters as policy maximizers---collapse under the triad's weight, as Downs' ideal gives way to group loyalty (Downs 1957; Achen and Bartels 2016). The triad's resilience suggests democracy adapts to these forces, not transcends them, a reality requiring pragmatic reckoning.

Cultural and institutional contexts mold the triad's expression, amplifying or tempering its effects. The U.S.'s individualistic, free-market culture views money as speech, clashing with egalitarian norms and driving distrust when racial identities polarize (Sandel 2012; Pew Research Center 2023). India's hierarchical diversity---caste, religion---pairs with lax regulation, letting money and information entrench identity blocs, sparking backlash when exclusion surfaces (Jaffrelot 2010; Vaishnav 2017). Germany's collectivist stability, rooted in post-war consensus, balances the triad through regulation, softening its edge while sustaining regional identities (Streeck 2014; Forschungsgruppe Wahlen 2021). China's Confucian emphasis on order aligns money and information with a unified identity, bolstered by centralized control (Bell 2015; Repnikova 2017). Russia's patrimonial tradition tolerates wealth concentration, fusing identity with state power, though shocks test this (Ledeneva 2013; Sakwa 2009). Culture and institutions thus govern the triad's visibility and stability, shaping its corrosiveness.

Policy implications demand tailored responses. For democracies, curbing the triad's excesses requires transparency and equity. In the U.S., tighter campaign funding rules could limit money's sway, aligning policy with voters over donors (Boatright 2015). India needs mandatory disclosure and spending caps post-bond rulings, restoring trust in a pluralistic system (The Hindu 2024). Germany could tighten private donation rules, enhancing its public funding model to reduce subtle biases (Transparency International 2018; McMenamin 2013). Information's role calls for regulation---curtailing microtargeting in the U.S., countering divisive content in India, moderating online fragmentation in Germany---to temper polarization (Sunstein 2017; Kumar and Singh 2014;

Barrett et al. 2021). Public funding boosts---across all three---could level competition, reducing reliance on identity-driven cash (Drutman 2015).

Non-democracies face a different challenge: diversifying legitimacy beyond the triad. China could expand social welfare, easing dependence on economic nationalism (Naughton 2018; Lardy 2021). Russia might shift resources into broader development---tech, infrastructure---diversifying from rents and patriotic identity, though elite resistance looms (Aslund 2019; Gurvich and Prilepskiy 2015). Both need mechanisms---like transparent audits or oversight---balancing control with resilience, a task complicated by dissent suppression (Li and Zhou 2024; Petrov 2019). Unlike democracies, where transparency drives reform, non-democracies must address hidden weaknesses internally, a delicate balance (Repnikova 2017). These reforms aim to manage the triad, not dismantle it, recognizing its constitutive role.

Theoretically, the triad refines established frameworks. Elite theory, per Mills, gains depth: money's structural power in democracies competes openly, while in non-democracies, it co-opts elites under state aegis (Mills 1956; Treisman 2020). Social identity theory extends to electoral dynamics, with money and information amplifying Tajfel and Turner's group loyalty across contexts (Tajfel and Turner 1979; Mason 2018; Chandra 2004). Media studies evolve, as Benkler's networked sphere becomes a tribal amplifier, necessitating visibility's role in legitimacy---transparency hastens corrosion in democracies, opacity delays it in autocracies (Benkler 2006; Bail et al. 2018; Frieden 2020). Modernization theory requires recalibration: the triad's persistence---racial in the U.S., caste in India, nationalist in China---defies issue-based evolution (Inglehart 1997; Lipset 1960). A refined model might cast the triad as a legitimacy multiplier---boosting power until overreach, with regime type setting the threshold (Parry 2020).

Broader implications stretch beyond these cases. Democracies must balance electoral freedom with equity, as the triad's unchecked flow---U.S. polarization, India's exclusion---threatens stability (Lessig 2011; Sridharan 2014). Non-democracies must diversify legitimacy, as reliance on money and identity wobbles under strain (Lardy 2021; Aslund 2019). Globally, the triad's feedback loop warrants scrutiny: money consolidates, identity motivates, and information amplifies, a cycle transcending borders (Drutman 2015; Kumar and Singh 2014). Civic education could foster cross-cutting identities, softening tribalism, while institutional design---proportional systems, power-sharing---might manage its tensions (Norris 2004; Kitschelt and Wilkinson 2007). The triad's permanence suggests democracy thrives not by overcoming these forces but by channeling them toward legitimate compromise.

The triad's dominance reframes polarization debates. In democracies, it's not an anomaly but an outcome---U.S. racial blocs, India's caste divides---intensified by money and information (Mason 2018; Jaffrelot 2010). Non-democracies mask it, unifying identity until cracks emerge (Treisman 2020). Health lies not in defeating the triad but in navigating it, through institutions legitimizing trade-offs (Lijphart 1999). Digital amplification heightens this challenge, demanding communication strategies engaging identity, not denying it (Sunstein 2017). The triad's implications thus span practice and theory, urging a shift from idealized visions---rational voters, deliberative spheres---to realistic accounts centering money, identity, and information as governance's core (Downs 1957; Habermas 1996). This perspective offers a path forward, grounding reform in the triad's inescapable reality. The stakes are high: unaddressed, the triad risks further democratic erosion, with declining faith in systems evident globally (Pew Research Center 2023).

A reorientation of democratic practice means adapting norms to channel identity positively, such as promoting inclusive patriotism (Norris 2004). Reforms like citizen assemblies or ranked-choice voting could inject thoughtful input and reward broader appeals, countering money and propaganda (Lessig 2011). Theoretical models should integrate elite theory, group theory, and media effects, treating the triad as fundamental to political outcomes (Achen and Bartels 2016). Democracy's resilience depends on confronting these challenges directly, ensuring money is equitable, identity inclusive, and information enlightening.

**8. Conclusion**

This perspective article illuminates the triad of money, identity, and information as a foundational force shaping modern democracies and beyond, weaving a narrative of power consolidation that transcends regime types. Across the United States, India, Germany, China, and Russia, the triad operates as a dynamic interplay: money fuels influence, identity binds loyalty, and information amplifies both, challenging the modernist vision of issue-based governance. The triad’s universality emerges from its adaptability. In the U.S., campaign financing, as detailed in Section 3, targets racial identities, as detailed in Section 4, disseminated through digital platforms, entrenching elite power and eroding trust (Gilens and Page 2014; Mason 2018; Bail et al. 2018; OpenSecrets 2024; PBS News 2024; NYU Stern 2021). India’s system, as noted in Section 3, channels significant funds through electoral bonds into caste and religious appeals, amplified by grassroots networks, consolidating dominance amid legitimacy strains (Kapur and Vaishnav 2018; Chandra 2004; Kumar and Singh 2014; The Hindu 2024; Ahmad 2024). Germany balances this with regulation, yet money and information sustain regional and national identities, with civic identity boosting turnout, subtly skewing representation (Elsässer et al. 2017; Arzheimer and Berning 2019; Mader and Schoen 2021). China and Russia invert it, using state wealth and controlled media to unify nationalist identities—Han Chinese pride, Eurasian patriotism—projecting stability until disruptions surface (Naughton 2018; Sakwa 2009; Treisman 2020; Foreign Affairs 2022; The Strategy Bridge 2019). Across these cases, the triad defies Downs' rational voter and Inglehart's post-materialist shift, entrenching tribalism over deliberation (Downs 1957; Inglehart 1997).

Legitimacy's fate reflects this duality. Democracies bear immediate corrosion—U.S. distrust, India's exclusion, Germany's tempered unease—as the triad's visibility exposes its distortions (Pew Research Center 2023; Gupta 2021; Forschungsgruppe Wahlen 2021). Non-democracies delay this, with money and information bolstering identity-based control, though risks loom beneath (Li and Zhou 2024; Gurvich and Prilepskiy 2015). This contribution refines theory: as established in Section 2, political economy and social identity theory underscore money's structural sway and group loyalty's role, and media studies frame information as an amplifier, challenging modernist linearity (Gilens and Page 2014; Tajfel and Turner 1979; Sunstein 2017; Kubina and von Sikorski 2021). The triad's feedback loop—money consolidating, identity motivating, information spreading—offers a lens to rethink power, not as a marketplace of ideas but as a contest of resources and tribes.

Implications span practice and scholarship. Democracies must curb the triad's excesses—reforming funding in the U.S., enhancing transparency in India, tightening rules in Germany—to restore trust (Boatright 2015; The Hindu 2024; Transparency International 2018). Non-democracies need diversified legitimacy—beyond China's prosperity or Russia's patriotism—to weather disruptions (Lardy 2021; Aslund 2019). Theoretically, the triad demands a shift from rationalist ideals to realistic models, centering money, identity, and information as governance's core (Achen and Bartels 2016). This resonates globally, as digital amplification and identity politics challenge stability worldwide (Kumar and Singh 2014; Mason 2018; Barrett et al. 2021).

Unresolved questions beckon further study. How will India's post-bond reforms reshape the triad's balance? Can Russia's resource-driven identity withstand prolonged strain? How will Germany's elections reflect regulated funding shifts? Can China's digital controls adapt to economic strain? Longitudinal analyses in democracies—tracking trust over time—and econometric models in non-democracies—probing stability's limits—could deepen insights (Vaishnav 2017; Treisman 2020). Cross-national comparisons—beyond these five—might test the triad's universality, while digital media's evolving role warrants scrutiny (Sunstein 2017; NYU Stern 2021). By blending empirical depth with comparative breadth, this article enriches discourse on power and legitimacy, urging scholars and policymakers to confront the triad's reality—not as a flaw to fix, but a force to manage—in an age of technological and social complexity.

**Statements and Declarations**

**Author Contributions:**

Both the authors contributed equally at all the stages of research leading to the submission of the manuscript.

**Funding Statement:**

This research did not receive any specific grant from funding agencies in the public, commercial, or not-for-profit sectors.

**Conflict of Interest:**

The authors declare no conflicts of interest related to this research.

**Data Availability:**

The manuscript does not report any new data.

**Software & AI Usage Statement:**

The authors made use of Chatgpt 4O and Grammarly to correct the language and the overall writing style of the manuscript. After using these tools, the authors reviewed and edited the content as needed and take(s) full responsibility for the content of the published article.